# A Low Profile Tunable Microwave Absorber based on Graphene Sandwich Structure and High Impedance Surface

Jin Zhang, Wei-Bing Lu, Zhen-Guo Liu, Hao Chen, Bian Wu, and Qi-Feng Liu

*Abstract*—In this paper, a low profile dynamically tunable microwave absorber is proposed, which consists of high impedance surface (HIS) and graphene sandwich structure (GSS). We theoretically demonstrate and experimentally verify that the proposed absorber can provide a dynamically tunable reflection range from larger than -3dB to less than -30 dB (corresponding to the absorption range from 50% to 99.9%) at operating frequency of 11.2GHz by external bias voltage. The entire thickness of this absorber is only 2.8mm, nearly one tenth of working wavelength. In addition, a modified equivalent circuit model is proposed to explicate its absorption mechanism. At last, we fabricate a prototype absorber, and measure its absorption in anechoic chamber. The experimental results agree well with the full wave simulation results. This work may provide a reference for design and fabrication of dynamically tunable microwave absorber based on large-scale graphene and may promote the actual applications of graphene at microwave frequency.

## I. INTRODUCTION

Graphene, a single layer of carbon atoms in a hexagonal lattice, has attracted a great deal of interests in electron devices due to its unique structure and outstanding physical characteristics [1]. Moreover, several works have been endeavored to design tunable electromagnetic components using graphene in recent years [2-10]. However, most of works mainly focus on tunable applications in THz band [2-6] and few works especially involves graphene-based tunable microwave absorbers (MA) [7-10]. Until now, graphene-based MA has been experimentally studied in several works [11-15]. Stacking graphene flakes based wideband absorber is achieved in [12]. In [13] Chen et al design and fabricate a MA using large-area multilayer-graphene based frequency selective surface. In [14], Yi et al propose a tunable MA based on patterned graphene. Although the working frequency of this MA can be directly changed by stacking different numbers of patterned graphene layers, lacking of dynamically tunable property and limitation of sample size restricts its more practical applications. Another switchable graphene-based X-band radar-absorbing surface is proposed in [15]. This absorbing surface can tune the amplitude of absorption by applying different bias voltage. It works as a configurable Salisbury screen so that the substrate layer's thickness is about a quarter of the operating wavelength, which does not satisfy the practical need of thin absorber. Therefore, there is still a great potential in extending the performance of graphene-based MA. In this work, we propose a low profile tunable MA using GSS and HIS, herein our proposed MA is referred as GSS-HISA in the following part. Unlike configurable Salisbury screen constrained by quarter-wavelength thickness, the entire thickness of our device can be reduced to nearly one tenth of working wavelength. The 3D full wave simulation result shows that by tuning the electrical bias voltage, our design can adjust the reflection amplitude from -33dB to -3dB (corresponding to absorption range from 99.9% to 50%). Then we explain the physical insight of this tunable absorber by using modified equivalent circuit model and multiple reflection theory. To verify the simulations, we fabricated an absorber sample with dimensions of 14cm×14cm and measured it by X-band standard lens antenna testing system. As a result, the measurement data shows a good agreement with that of equivalent circuit model and 3D full wave simulation results.

## II. STRUCTURE DESIGN

As shown in Fig.1 (a) and (b), the GSS-HISA comprises of five components: a GSS on the top, a spacer layer, a MFSS layer in the middle, a dielectric slab as the substrate layer, and a ground plate at the bottom. The GSS part consists of two monolayer graphene layers and a piece of paper tissue soaked with electrolyte between two graphene layers. The HIS is a planar-periodic structure and each unit cell comprises an identical square loop. The relative permittivity of spacer layer $\varepsilon_{spa}$=1.05 and the relative permittivity of substrate $\varepsilon_{sub}$=1.05. $t_{ele}$, $t_{spa}$ and $t_{sub}$, which are correspond to the thickness of electrolyte tissue, spacer layer and substrate, are equal to 50μm, 1.2mm and 1.575mm, respectively.

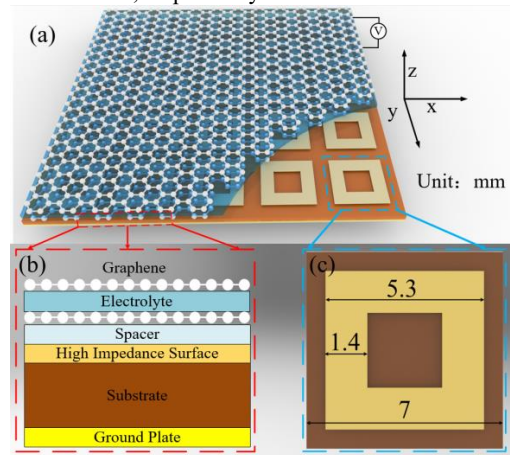

Fig.1 (a) Three-dimensional structure of the proposed GSS-MFSSA. (b) Side view of the structure comprised of five components: GSS layer, spacer layer, MFSS layer, substrate layer and ground plane. (c) Top view of a unit cell of metal frequency selective surface layer.

It has been studied in the previous works that the surface conductivity of graphene $\sigma_{gra}$ is highly dependent on the working frequency and chemical potential (or Fermi energy) [16]. When there is no external magnetic field, the local surface conductivity is isotro-

Manuscript received xxxxx, 2018, revised xxxxx; accepted xxxxx. Date of publication xxxxx; date of current version xxxxx. This work is supported by the National Natural Science Foundation of China (NSFC) under grant number 61671150, 61671147, 61771360 and Six talent peaks project in Jiangsu Province XCL-004. (Corresponding author: Wei-Bing Lu, e-mail: wblu@seu.edu.cn)

J. Zhang, W. B. Lu, Z. G. Liu and H. Chen are with State Key Lab of Millimeter waves, School of Information Science and Engineering, Southeast University, Nanjing 210096, China.

B. Wu is with the National Key Laboratory of Antennas and Microwave Technology, Xidian University, Xi'an 710071, China.

Q. F. Liu is with the Science and Technology on Electromagnetic Compatibility Laboratory, China Ship Development and Design Centre, Wuhan 430000, China.

Color versions of one or more of the figures in this communication are available online at http://ieeexplore.ieee.org.

Digital Object Identifier 10.1109/TAP.2016.xxx

pic, which can be calculated directly by the Kubo formula [17]. So the sheet impedance of monolayer graphene can be calculated as:

$$R_{gra} + jX_{gra} = \frac{1}{\sigma_{gra}} \approx R_{gra} \quad (1)$$

Note that in microwave frequency range, the imaginary part of the sheet impedance of graphene ($X_{gra}$) can be ignored and monolayer graphene can be regarded as a resistive sheet without dispersion property. According to parallel theory, the total equivalent impedance of GSS can be represented by only two identical graphene sheet resistances in parallel:

$$R_{GSS} = R_{gra} \parallel R_{gra} \quad (2)$$

In our GSS, the conductivity of electrolyte is 0.22S/m and $R_e$ (the surface resistance of electrolyte layer) is equal to 27000Ω. Based on the transmission line theory, an equivalent circuit analog model for the GSS-HISA is established in Fig 2. The GSS and HIS can be described by the equivalent impedances $R_{GSS}$ and $Z_{HIS}$ respectively.

$$Z_{HIS} = j\omega L_m + \frac{1}{j\omega C_m} \quad (3)$$

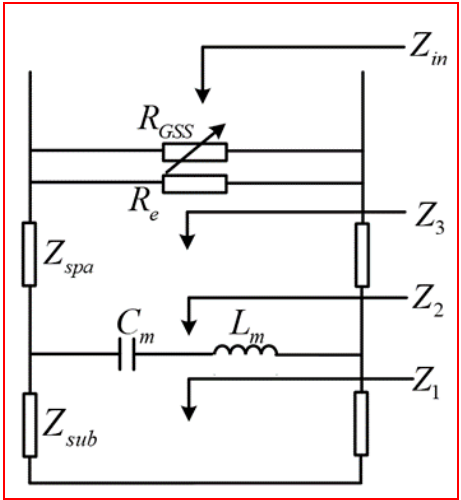

Fig.2 The equivalent circuit model of the graphene-MFSS composite absorber based on transmission line theory. The GSS is represented with two tunable resistances in parallel, the MFSS is equivalent to an LC series circuit and the metal ground plate is shown with a short circuit.

The original derivation of equivalent circuit method for HIS started from the analysis of infinite grating strips. The infinite grating metallic strips array is represented by a frequency-dependent inductor $L$ or capacitor $C$. As for metallic square loop array, considering as the combination of grating strips distributed along two orthogonal directions, it could be modeled as series $LC$ circuit [18].

In addition, the spacer layer and the substrate layer, regarded as transmission lines in our model, can be designed either inductive reactance or capacitive reactance. On the condition of normal incidence, the effective circuit parameters for TE polarization (the electric field propagates along x-axis) are derived as follows [19]:

$$Z_1 = jZ_{sub} \tan(k_{sub} t_{sub}) \quad (4)$$

$$Z_2 = \frac{Z_{HIS} \cdot Z_1}{Z_{HIS} + Z_1} \quad (5)$$

$$Z_3 = Z_{spa} \frac{Z_2 + jZ_{spa} \tan(k_{spa} t_{spa})}{Z_{spa} + jZ_2 \tan(k_{spa} t_{spa})} \quad (6)$$

$$Z_{in} = R_{GSS} \parallel R_e \parallel Z_3 \quad (7)$$

where $Z_{spa} = \omega\mu_0 / k_{spa}$ and $Z_{sub} = \omega\mu_0 / k_{sub}$ are the characteristic impedance of spacer layer and substrate layer respectively. $k_{spa} = \omega\sqrt{\varepsilon_0 \varepsilon_{spa} \mu_0}$ and $k_{sub} = \omega\sqrt{\varepsilon_0 \varepsilon_{sub} \mu_0}$ are the propagation constants of different layers along z direction.

When keeping other geometric parameters unchanged, it is worth mentioning that $Z_{in}$ is a function of $R_{GSS}$ and frequency $f$. Therefore, it can be expressed as $Z_{in}(R_{GSS}, f)$. According to transmission line theory, when a normal incident wave impinges on the GSS-HISA, the reflection coefficient $S_{11}$ can be derived from:

$$S_{11} = \frac{Z_{in}(R_{GSS}, f) - \eta_0}{Z_{in}(R_{GSS}, f) + \eta_0} \quad (8)$$

where $\eta_0$ is the characteristic impedance of free space. In the case of a conductor-backed absorber, because the transmission is blocked by the ground plate at the bottom, the absorptivity of the proposed MA can be represented by reflection coefficient, $A(f) = 1 - |S_{11}|^2$. Therefore, the conditions for perfect absorption($S_{11}$=0) are given as:

$$Re[Z_{in}(R_{GSS}, f_0)] = \eta_0 \approx 377\Omega \quad (9)$$

$$Im[Z_{in}(R_{GSS}, f_0)] = 0 \quad (10)$$

We set an optimized frequency $f_0$ firstly and then adjust the value of $R_{GSS}$ to satisfy these two conditions. For a given frequency, it is obvious that $R_{GSS}$ is a key parameter and has significant influence in perfect absorption performance.

### III. SIMULATION AND ANALYSIS

We consider that a TE polarized plane wave propagating along -z direction impinges on the GSS-HISA. Commercial software CST2015 is employed to simulate the reflection coefficient $S_{11}$ of this structure. In addition, numerical calculations based on equivalent circuit model are performed as a contrast. The comparison results from these two methods under different values of $R_{GSS}$ are illustrated in Fig.3 (a). The range of $R_{GSS}$ is from 1200Ω to 300Ω, which derived from our previous works[20]. Fig.3 (a) shows a good agree-

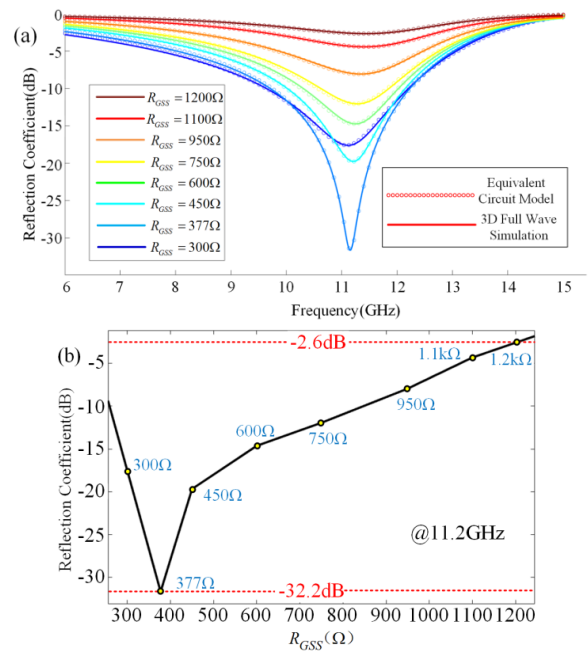

Fig.3 (a) Reflection coefficients of GSS-MFSSA versus frequency under different values of $R_{GSS}$. Solid line represents the 3D full wave Simulation results and circular line means the equivalent circuit results. (b) Simulated reflection coefficient of GSS-MFSSA versus $R_{GSS}$ at 11.2GHz.

ment between 3D full wave simulations and equivalent circuit model calculation results, which proves the validity of the equivalent circuit model. In addition, it is obviously illustrated that by increasing the value of $R_{GSS}$ from 300Ω to 1200Ω, the magnitude of reflection coefficient changes dramatically. In order to clarify their relations, Fig. 3 (b) shows the reflection coefficient as a function of $R_{GSS}$ at operating frequency 11.2GHz. When setting $R_{GSS}$ is equal to 1200Ω, the simulated reflection coefficient at 11.2GHz is about -2.6 dB, which means that the GSS-HISA reflects most of incident wave and we name this state "the reflection mode". As the value of $R_{GSS}$ decreases from 1200Ω to 377Ω gradually, the reflection coefficient reduces from -2.6dB to -32.2dB dramatically. Notably, the reflection coefficient reaches its minimization (about -32.2dB) as the value of $R_{GSS}$ changes to 377Ω. In this case, almost all of incident wave is absorbed in the GSS-MFSSA and we denominate this case as "the absorption mode". Keeping decreasing the value of $R_{GSS}$ from 377Ω to 300Ω, the reflection coefficient is no longer decline. On the contrary, it rises up from -32.2dB to -18dB at 11.2GHz. In a word, according to simulation results in Fig.3 (b), it has been verified that the reflection coefficient of the GSS-MFSSA can be tuned from -2.6 dB to -32.2dB by changing the value of $R_{GSS}$ which shows that our proposed structure has a tunable wide-range of absorption.

We next adopt the multiple interference theory to analyze the absorption of the MA quantitatively [21,22]. The interference model for the absorber with GSS and FSS is depicted in Fig. 4. This model contains three interfaces, the GSS interface, the FSS interface and copper film. The GSS and FSS acts as a part of the reflection surface that can reflect/transmit a part of the incident wave. The copper film works as a perfect reflector with a reflectance of $\tilde{r}_{33} = -1$, thus the transmission through the MA is zero.

face 2 and interface 1 with coefficients of $\tilde{r}_{32}=r_{32}e^{j\phi_{32}}$, $\tilde{r}_{21}=r_{21}e^{j\phi_{21}}$, $\tilde{t}_{32}=t_{32}e^{j\theta_{32}}$ and $\tilde{t}_{21}=t_{21}e^{j\theta_{21}}$ respectively. Similar to the light propagation in a stratified media, the overall reflection is the superposition

$$\tilde{R} = \tilde{r}_{12} + \tilde{t}_{12}\tilde{r}_g\tilde{t}_{21}e^{j2\beta_1} + \tilde{t}_{12}\tilde{r}_g^2\tilde{r}_{21}\tilde{t}_{21}e^{j4\beta_1} + \tilde{t}_{12}\tilde{r}_g^3\tilde{r}_{21}^2\tilde{t}_{21}e^{j6\beta_1} + \cdots$$

$$= \frac{\tilde{r}_{12} - (\tilde{r}_{12}\tilde{r}_{21} - \tilde{t}_{12}\tilde{t}_{21})\tilde{r}_g e^{j2\beta_1}}{1 - \tilde{r}_g\tilde{r}_{21}e^{j2\beta_1}} = \frac{\tilde{r}_{12} - \tilde{\alpha}}{1 - \tilde{r}_g\tilde{r}_{21}e^{j2\beta_1}} \quad (11)$$

$$\tilde{r}_g = \tilde{r}_{22} + \tilde{t}_{23}\tilde{r}_{33}\tilde{t}_{32}e^{j2\beta_2} + \tilde{t}_{23}\tilde{r}_{33}^2\tilde{r}_{32}\tilde{t}_{32}e^{j4\beta_2}$$
$$+ \tilde{t}_{23}\tilde{r}_{33}^3\tilde{r}_{32}^2\tilde{t}_{32}e^{j6\beta_2} + \tilde{t}_{23}\tilde{r}_{33}^4\tilde{r}_{32}^3\tilde{t}_{32}e^{j8\beta_2} + \cdots$$

$$= \frac{\tilde{r}_{22} + (\tilde{t}_{23}\tilde{t}_{32} - \tilde{r}_{22}\tilde{r}_{32})\tilde{r}_{33}e^{j2\beta_2}}{1 - \tilde{r}_{33}\tilde{r}_{32}e^{j2\beta_2}} \quad (12)$$

of the multiple reflections at the GSS interface, it can be expressed in formula (11), where $\beta_{1,2}=\sqrt{\varepsilon_{pet}}k_0h_{1,2}$ is the propagation phase in the air and spacer repectively, $k_0$ is the free space wavenumber and $\tilde{\alpha} = (\tilde{r}_{12}\tilde{r}_{21} - \tilde{t}_{12}\tilde{t}_{21})\tilde{r}_g e^{j2\beta_1}$. To attain a nearly perfect absorption at the working frequency, which means the reflectance is almost zero, the overall reflection coefficient should satisfy the following conditions:

$$|\tilde{r}_{12}| = |\tilde{\alpha}| \quad (13)$$

$$\varphi(\tilde{r}_{12}) - \varphi(\tilde{\alpha}) = 2n\pi, |n| = 0,1,2,\ldots \quad (14)$$

As depicted in Fig.5, on one hand, the amplitudes of $\tilde{r}_{12}$ and $\tilde{\alpha}$ are approximately equal in 2GHz~16GHz, satisfying the amplitude condition. On the other hand, the phase difference $\varphi(\tilde{r}_{12}) - \varphi(\tilde{\alpha})$ is equal to 0° only at 11.2 GHz, which perfectly fulfills the phase condition. In a word, the amplitude and phase conditions are simultaneously satisfied at 11.2 GHz and hence a perfect absorption is achieved. In a word, the physical mechanism of tunable absorption

At the interface1, the incident wave is partially reflected back to air with a reflection coefficient $\tilde{r}_{12}=r_{12}e^{j\phi_{12}}$ and partially transmitted into spacer with a transmission coefficient $\tilde{t}_{12}=t_{12}e^{j\theta_{12}}$. The latter continues to propagate until it reaches the interface 2, with a complex propagation phase $\beta_1$. It is partially reflected back to spacer with a reflection coefficient $\tilde{r}_{22}=r_{22}e^{j\phi_{22}}$ and transmitted into substrate with a transmission coefficient $\tilde{t}_{23}=t_{23}e^{j\theta_{23}}$. After the reflection at the interface 3 and addition of another propagation phase $\beta_2$, partial reflection and transmission occur again at the inter-can be described as below. When the incident EM waves impinge normally on absorber's surface, some EM waves are transmitted through GSS layer and HIS layer while others are reflected directly. Transmitted EM waves then are reflected back by the metallic ground plate and come out of the structure with a phase difference. This process is repeated again and again, and these partial reflections on the top surface of absorber destructively interfere with each other, leading to consumption in the end. GSS layer is the mostly responsible to match the impedance of the absorber structure to the free space impedance while HIS layer is designed to reduce the whole thickness of absorber. Fig.6 shows that both these two absorber can realize tunable microwave absorption by changing the value of $R_{GSS}$. It is worth mentioning that the thick-ness of previous graphene based Salisbury screen must be fixed as a quarter of wavelength, which means 6.7mm in case of 11.2GHz. However, owing to the HIS, the thickness of this work is only 2.8mm (nearly one tenth of working wavelength), which makes our absorber low-profile and affords a further wide field of practical application.

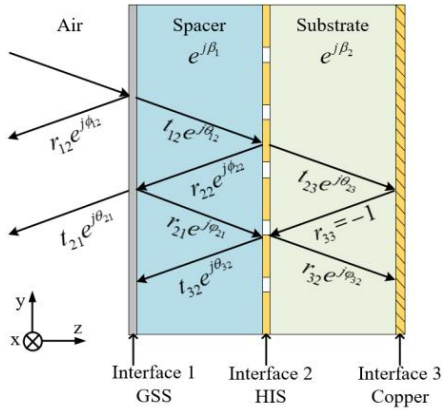

Fig. 4. The interference model of the absorber and associated variables.

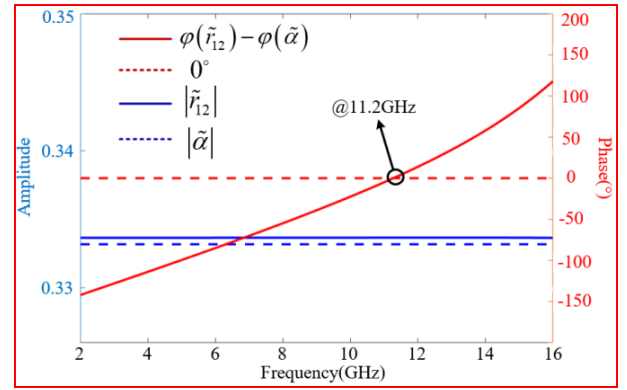

Fig.5. Phase of $\varphi(\tilde{r}_{12}) - \varphi(\tilde{\alpha})$ and amplitude of $\tilde{r}_{12}$, $\tilde{\alpha}$.

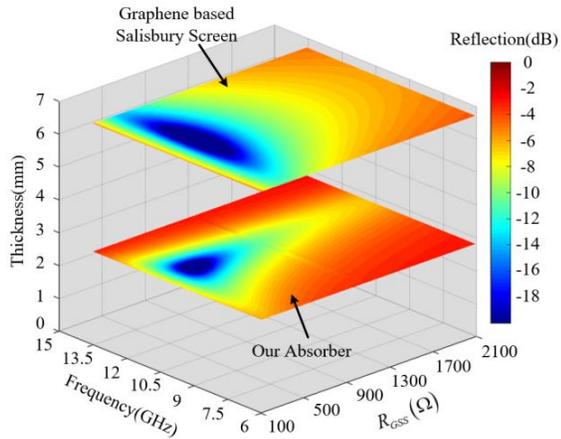

Fig.6 Comparison of calculated reflection coefficient between our absorber and previous graphene based Salisbury screen.

In the analysis above, the reflection coefficient of the GSS-MFSSA has been investigated under normal incidence for x-polarization. Nevertheless, it is necessary to discuss the situations about oblique incidence. Thus, as shown in Fig.7, we simulate the effect of incident angles for both TE and TM configuration. It illuminate that the reflection coefficient obviously becomes worse and the resonant frequency shifts to high frequency region as the incident angle increases for both TE and TM polarization. This is because that the magnetic field component becomes small with the increasing of the incident angle, and cannot efficiently excite the resonance at metal frequency selective surface. However, the reflection coefficient is still remains below -15dB for the incident angle up to 45°. In addition, compared Fig.7(a) with Fig.7(b), the difference of polarization mode has slight influence in oblique incidence. Therefore, owing to the high-degree symmetry of each unit cell, the GSS-MFSSA exhibits a good polarization-insensitive performance.

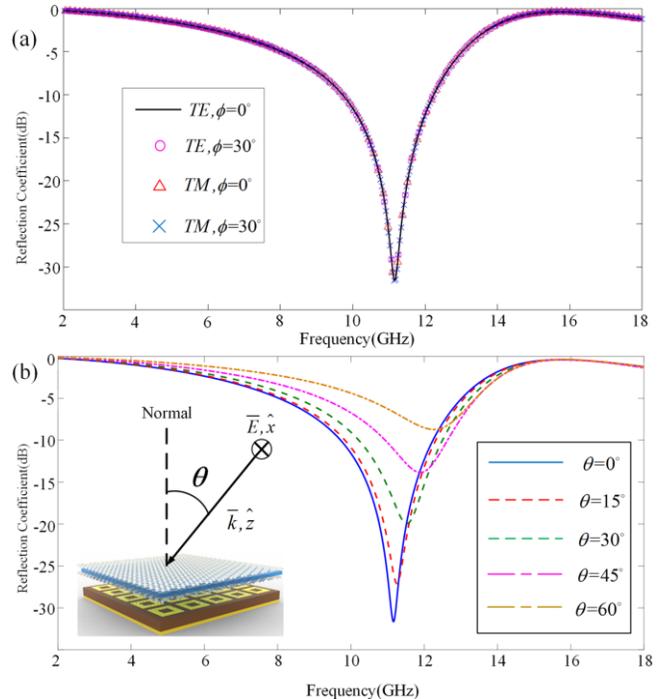

Fig.7 Simulated reflection coefficient for different incidence angles for (a) TE polarized wave (b) TE polarized wave.

IV. EXPERIMENTAL RESULTS

The ability to synthesize and transfer high-quality and large-scale graphene enables us to realize our proposed graphene-based MA. The detailed progress to fabricate graphene films has been introduced in Ref. [13]. Two thin copper stripes at the edges of the graphene film served as contact metals to apply voltage to the graphene electrodes, as shown in Fig.8(a). According to [15], we know that this graphene sandwich structure resembles super capacitors and the graphene layers connected to negative voltage gets elec- tron-doped, whereas the other one gets hole-doped. Therefore, the sheet resistance of two graphene layers in sandwich structure ought to be tuned by applying bias voltage simultaneously. A 20×20 unit cell HIS sample, as shown in Fig.8(b), has been fabricated on a 140mm×140mm planar sheet of 1.575mm thick Rogers 5880 series through standard printed circuit board (PCB) technology. Then we fixed the graphene sandwich structure and MFSS sample together. At last, we put a 1.2mm thick polyethylene as a spacer layer in the middle of the whole structure. The final fabricated prototype of graphene-MFSS composite absorber is illustrated in Fig. 8(c).

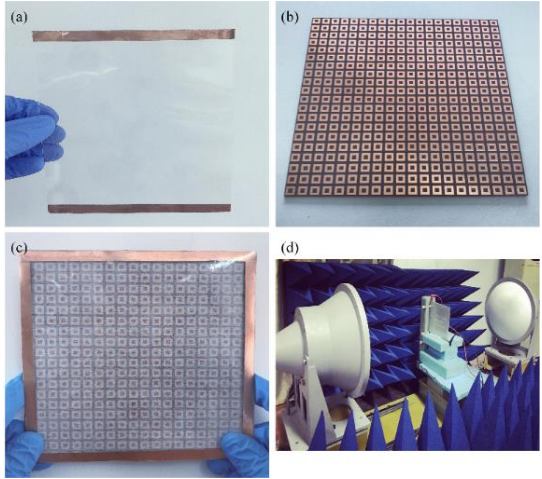

Fig.8 (a) Photograph of a monolayer graphene film with metallic electrodes (b) Photograph of MFSS layer and substrate layer fabricated by using PCB technology. (c) Photograph of the fabricated GSS-MFSSA sample. (d) Configuration of the measurement setup.

As shown in Fig. 8(d), our experimental measurement is carried out in a small microwave chamber. Keithley 2400 source measure unit works as a DC power supply connecting to the positive and negative copper electrode of graphene layer. A standard gain X-band meniscus lens antennas connected to a vector network analyzer (Keysight N5222A) serves as the transmitter and receiver during the reflection spectra measurement. The measured and simulated reflection coefficients of the GSS-HISA applied different bias voltage are illustrated in Fig. 8. The measured resonances occurred at 11.2GHz are roughly consistent with the simulations and equivalent-circuit calculation results. However, slight difference between measured results and simulated results exists at the resonance point, which leads to a blue shift of working frequency. This frequency discrepancy may be caused by the fabrication tolerance. Besides that, the wires that are used to connect voltage source to graphene layer may interfere the measured results slightly as well. Despite these discrepancies, it is obvious that the experimental results show a very good agreement with the results of 3D full wave simulation and the dynamically tunability of our designed GSS-HISA can be validated.

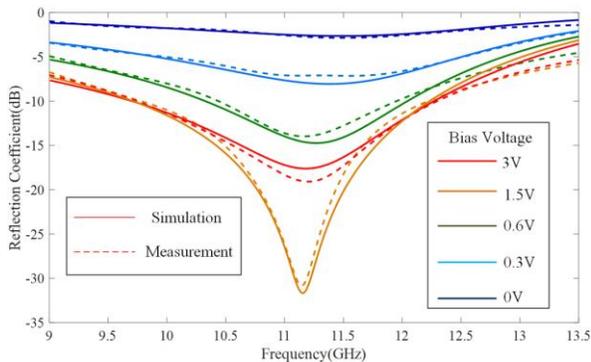

Fig.9 Measured and simulated reflection coefficient of the graphene-FSS component absorber under different bias voltage from 0V to 3V.

It is worth mentioning that $R_{GSS}$ is a key parameter during the simulation process and it works as a bridge to connecting reflection coefficient with bias voltage. This property can be used as a reference for designing microwave component based on GSS in the future.

## V. CONCLUSION

In summary, a design of low profile dynamically tunable microwave absorber is proposed in this paper, which consists of GSS and HIS. Unlike configurable Salisbury screen constrained by quarter-wave thickness, the entire thickness of our device can be set as nearly one tenth of working wavelength due to high impedance surface structure. The 3D full wave simulation and experimental measurement demonstrated that the reflection of our device can be tuned from nearly -30dB to large than -3dB by changing the surface resistance of graphene. Equivalent circuit model and multiple interference theory are introduced to explain the absorption mechanism of the proposed absorber. This work provides a reference for tunable ultrathin microwave absorber based on monolayer graphene and has actual significance to overcome the application difficulties of graphene structure at microwave band.